\documentclass[12pt,twoside]{article}
\usepackage{fleqn,espcrc1}
\usepackage{graphicx}

\newcommand{\zaa}{Astron. Astrophys.~}
\newcommand{\zapj}{Astrophys. J.~}
\newcommand{\znp}{Nucl.~Phys.~}

\newcommand{\zMNRAS}{Mon. Not. R. Astron. Soc.}

\begin{document}

\title{The Impact of the NACRE Compilation on the Big Bang Nucleosynthesis}

\author{Elisabeth Vangioni-Flam\address[IAP]{Institut d'Astrophysique de
Paris, 98$^{bis}$ Bd Arago 75014 Paris, France},
Alain Coc\address[CSNSM]{
CSNSM, IN2P3-CNRS-UPS, B\^atiment 104, 
91405 Orsay Campus, France} and
Michel Cass\'e\address[SAP]{
SAp, DAPNIA, DSM, CEA, Orme des Merisiers, 91191 Gif sur Yvette CEDEX
France} 
\addressmark[IAP]}

\maketitle

\begin{abstract}
We update the Big Bang Nucleosynthesis (BBN) calculations on the basis of
the recent NACRE compilation of reaction rates.
We estimate the uncertainties related to the nuclear reaction rates on the
abundances of $D$, ${^3}He$, ${^4}He$, ${^6}Li$, ${^7}Li$, ${^9}Be$, 
${^{10}}B$ and ${^{11}}B$ of cosmological and astrophysical interest.
We use lithium as the main indicator of
the baryon density of the Universe, rather than deuterium. 
\end{abstract}

\section{Introduction}

Recently, the new NACRE compilation of thermonuclear rates\cite{NACRE} has
been published and observations of light isotopes have flourished. It is
thus timely to reassess the determination of the baryonic density of
the Universe in the light of advances in nuclear physics and astronomical
observations.
To calculate primordial abundances up to $B$, the BBN network is constituted
by about 60 reactions\cite{Van00b} out of which, 22 are included in the NACRE
compilation. The main improvements with respect to former compilations 
concern a better traceability to the source of
nuclear data and the availability of upper and lower rate limits.
The primordial abundances of the light elements $D$, ${^{3,4}}He$ and
${^7}Li$ are governed by the expansion rate of the Universe and the cooling
it induces. Under the classical assumptions, these abundances depend only on
the baryon to photon ratio $\eta$, related to the baryonic parameter by
$\eta_{10} = 273.\Omega_B.h{^2}$ with $h = H/100$ km/s.Mpc.

\section{Discussion}

\begin{figure}[t]
\begin{center}
\includegraphics[height=10cm]{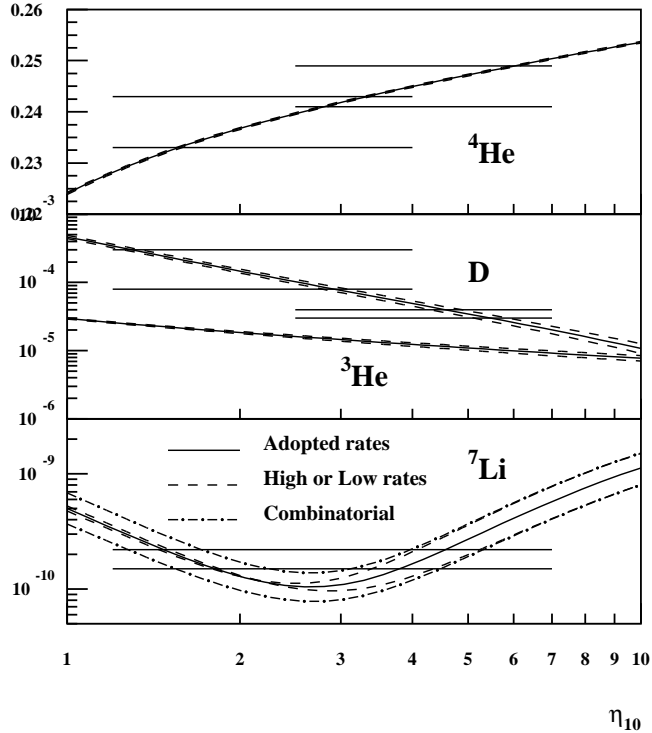}
\caption{Theoretical primordial abundances  of ${^4}He$ (by mass), $D$, 
${^3}He$ and ${^7}Li$ (by number) vs the baryon/photon ratio, $\eta$, 
using the reaction rates from the NACRE compilation\protect\cite{NACRE}. 
Solid lines: using recommended values of the reaction rates.
Dashed lines: using high or low  values of the reaction 
rates\protect\cite{Van00b} with possible error compensation.
Dash-dotted lines: most extreme results following all combination of 
high and low rates.
Horizontal lines indicate the error bands related to 
different observations.
}
\end{center}
\end{figure}

$D/H$ is measured in three astrophysical and/or cosmological sites : local
interstellar medium : 0.5 to 2. $10{^{-5}}$ \cite{Lin95,Vid00}, protosolar
nebula : 3.$\pm$0.3 $10{^{-5}}$ \cite{Dro99}, absorbing clouds on the
sightlines of quasars : 3. to 4. $10{^{-5}}$, \cite{Tyt00} and
0.8 to 3. $10{^{-4}}$ \cite{Web97}.
The primordial abundance of ${^4}He$, $Y_p$ is measured in low metallicity
extragalactic HII regions and in HII regions in the Small Magellanic Cloud.
Two ranges emerge: $Y_p$ = 0.238  $\pm$ 0.004 \cite{Fie98,Pei00}
and  $Y_p$ = 0.245  $\pm$ 0.004 \cite{Izo98}.
Determination of $Li$ in halo stars\cite{Spi96,Bon97,Rya99} indicate that
the Spite plateau is exceptionally thin ($<$ 0.05~dex).
We adopt the following range: 1.5 $10{^{-10}}$ $<$ ${^7}Li/H$ $<$ 2.2
$10{^{-10}}$, taking into account a maximum destruction of 0.1~dex.
The ${^6}Li$ abundance has also been determined in two halo
stars\cite{Hob97,Smi98,Cay99}.
We confirm that BBN calculated abundances of ${^9}Be$, ${^{10}}B$ and
${^{11}}B$ are negligible with respect the measured ones in the more metal
poor stars.
The uncertainty on the $D(\alpha,\gamma){^6}Li$ rate, and hence on of ${^6}Li$
production, is high. 
At the upper limit, it could lead to a primordial ${^6}Li$ abundance within
reach of future observations, to be used as a new cosmological constraint. 
Anyway, 
the bulk of the abundances of these elements can be explained in term of 
spallation of fast carbon and oxygen in the early Galaxy\cite{Van00a}.
Considering ${^7}Li$ observational constraints (see Fig.~1), in the most
favorable case where errors compensate\cite{Van00b} two possible ranges
emerge: $i)$ $1.5 < \eta_{10} < 1.9$ corresponding to
$0.013 < \Omega_B < 0.019$ ($h_{100} = 0.65$) in good concordance with
the error boxes related to the observed high $D$ and low ${^4}He$
$ii)$ $3.3 < \eta_{10} < 5.1$ corresponding to $0.029 < \Omega_B < 0.045$
in fair agreement with  a low $D/H$ and a high ${^4}He$. 
In the most pessimistic case, where errors are maximized (considering all
combinations of high and low NACRE rates, see Fig.~1) 
the $\eta_{10}$ ranges are wider (1.2 to 2) and (3 to 5.1).
Comparing the baryonic density to that of luminous matter 
($0.002 < \Omega_L < 0.004$ \cite{Sal99}) in the Universe 
shows the necessity of baryonic dark matter.
The observations of the Lyman alpha forest clouds between the redshifts 0 
and 5, lead to a corresponding  $\Omega_B = 0.03 \pm 0.01$, taking into 
account the uncertainty related to ionized hydrogen\cite{Rie98}.
This value is thought to reflect the bulk of the baryons at large scale.

\section {Conclusion}

Big bang nucleosynthesis
deserves permanent care since it gives access to the baryon density which
is a key cosmological parameter.
This work has been aimed at integrating the last development in both fields
of nuclear physics and observational abundance determination of light
elements.
The update of the reaction rates of the BBN using the NACRE
compilation has been made.
${^6}Li$ is affected by the large uncertainty of the
$D(\alpha,\gamma){^6}Li$ reaction. A refined measurement
of this reaction is of great cosmological interest.
Owing to the high observational reliability of the halo star ${^7}Li$
abundance data with respect to the $D$ data available which are both rare
and debated, we choose it as the leading baryometer.
Due to the valley shape of the Li curve, we get two possible $\eta$ ranges
which are consistent respectively with a high $D$ and low ${^4}He$ observed
values and with a low $D$ and high ${^4}He$ observed values. 
At present, none of these solutions can be excluded.



\begin{thebibliography}{30}

\bibitem{NACRE} C.~Angulo, M.~Arnould, M.~Rayet et al., \znp\ A656
(1999) 3.
\bibitem{Van00b} E.~Vangioni-Flam, A.~Coc \& M.~Cass\'e, \zaa\ 360 (2000) 15.
\bibitem{Lin95} J.~Linsky et al., \zapj\ 451 (1995) 335.
\bibitem{Vid00} A.~Vidal-Madjar, in 'The light Elements and their Evolution',
IAU Symp. 198, ASP Conf. Series., Edts L. da Silva, R. de Meideros and 
M. Spite, (2000) in press
\bibitem{Dro99} A.~Drouart, B.~Dubrulle, D.~Gautier \& F.~Robert, Icarus
140 (1999) 129.
\bibitem{Tyt00} D.~Tytler et al., in 'Nobel Symposium 109: Particle Physics
and the Universe', Physica Scripta T85 (2000) 12.
\bibitem{Web97} J.K.~Webb et al.,  Nature 383 (1997) 250.
\bibitem{Fie98} B.~Fields \& K.~Olive, \zapj\ 506 (1998) 177.
\bibitem{Pei00} M.~Peimbert \& A.~Peimbert, in 'The light Elements
and their Evolution', IAU Symp. 198, ASP Conf. Series.  Edts L. da Silva,
R. de Meideros and M. Spite, (2000) in press.
\bibitem{Izo98} Y.I.~Izotov \& T.X.~Thuan, \zapj\ 500 (1998) 1888.
\bibitem{Spi96} M.~Spite, P.~Francois, P.E.~Nissen \& F.~Spite, 
\zaa\ 307 (1996) 172.
\bibitem{Bon97} P.~Bonifacio \& P.~Molaro, \zMNRAS\ 285 (1997) 847.
\bibitem{Rya99} S.G.~Ryan, J.E.~Norris \& T.C.~Beers, \zapj\ 523 (1999) 654.
\bibitem{Hob97} L.M.~Hobbs \& J.A.~Thorburn, \zapj\ 491 (1997) 772.
\bibitem{Smi98} V.V.~Smith, D.L.~Lambert \& P.E.~Nissen, \zapj\ 506 (1998) 
405.
\bibitem{Cay99} R.~Cayrel et al., \zaa\ 343 (1999) 923.
\bibitem{Van00a} E.~Vangioni-Flam, M.~Cass\'e \& J.~Audouze,
Physics Report 333-334 (2000) 365.
\bibitem{Sal99} P.~Salucci \& M.~Persic,  \zMNRAS\ 309 (1999) 923.
\bibitem{Rie98} R.~Riedeger, P.~Petitjean \& J.P.~Mucket, \zaa\ 329 (1998) 30.

\end{thebibliography}
\end{document}